\RequirePackage{relsize,fix-cm}
\documentclass[english,aps,prl,reprint,twocolumn,amsmath,amssymb]{revtex4-2}

\usepackage{newtxtext,newtxmath}

\usepackage{graphicx}

\usepackage[dvipsnames,svgnames,x11names,hyperref]{xcolor}
\usepackage{hyperref}
\hypersetup{ 
    colorlinks=true,
    linkcolor = NavyBlue,
    urlcolor  = NavyBlue,
    citecolor = NavyBlue
}

\usepackage{bbm,bm,dsfont}
\usepackage{physics,braket,siunitx,slashed}
\renewcommand\vec{\bm} 

\newcommand{\valley}{v}
\newcommand{\fermi}{{{\scriptscriptstyle {\rm F}}}}

\newcommand{\pdagger}{{\phantom{\dagger}}}
\newcommand{\figref}[1]{Fig.~\ref{fig:#1}}

\newcommand{\pmbracket}{\mathbin{\vcenter{\hbox{%
  \oalign{\hfil\mkern-3mu$\scriptstyle+$\mkern-3mu\hfil\cr
          \noalign{\kern-.5ex}
          $\scriptscriptstyle({-})$\cr}
}}}}

\begin{document}
\relscale{1.06}

\title{Electrically Tunable Flat Bands and Magnetism in Twisted Bilayer Graphene}

\author{T. M. R. Wolf}
\author{J. L. Lado}
\author{G. Blatter}
\author{O. Zilberberg}
\affiliation{\mbox{Institute for Theoretical Physics, ETH Zurich, 8093 Zurich, Switzerland} \linebreak[4]}

\begin{abstract}
    Twisted graphene bilayers provide a versatile platform to engineer
    metamaterials with novel emergent properties by exploiting the resulting
    geometric moir\'{e} superlattice.  Such superlattices are known to host
    bulk valley currents at tiny angles ($\alpha\approx 0.3 ^\circ$) and flat
    bands at magic angles ($\alpha \approx 1^\circ$).  We show that tuning the
    twist angle to $\alpha^*\approx 0.8^\circ$ generates flat bands away from
    charge neutrality with a triangular superlattice periodicity.  When doped
    with $\pm 6$ electrons per moir\'e cell, these bands are half-filled and
    electronic interactions produce a symmetry-broken ground state (Stoner
    instability) with spin-polarized regions that order ferromagnetically.
    Application of an interlayer electric field breaks inversion symmetry and
    introduces valley-dependent dispersion that quenches the magnetic order.
    With these results, we propose a solid-state platform that realizes
    electrically tunable strong correlations.
\end{abstract}

\date{\today}

\maketitle

Controllably engineering quantum states of matter is one of the leading goals
of modern physics. This basic idea has been realized in a plethora of platforms
ranging from cold-atom setups~\cite{Bloch2008, Mazurenko2017, Uehlinger2013} to
atom-by-atom deposited solids~\cite{Toskovic2016, Loth2012}. In recent years,
the discovery of graphene has opened numerous new avenues
\cite{CastroNeto2009}, including the possibility of stacking two-dimensional
crystals and forming so-called `van der Waals' materials~\cite{Geim2013}, which
allow to engineer exotic states~\cite{Geim2013, Wolf2018, Cao2018,
Cao2018super, Rickhaus2018}. Tuning the relative angle $\alpha$ between
graphene layers \cite{RibeiroPalau2018} and applying electric potentials $V$
across the layers \cite{Castro2007, Rickhaus2018} have played central roles in
purposely designing the physical properties of such systems.  In this Letter, we
combine these two ideas in a new parameter regime: (i) we identify an angle
$\alpha$ that generates flat bands with strong correlations leading to a
magnetic instability, while (ii) an electric bias $V$ across the layers
reintroduces dispersion and thus allows to dynamically tune the magnetic
response of the bilayer system.

\begin{figure}[!t]
\includegraphics[width=3.5in]{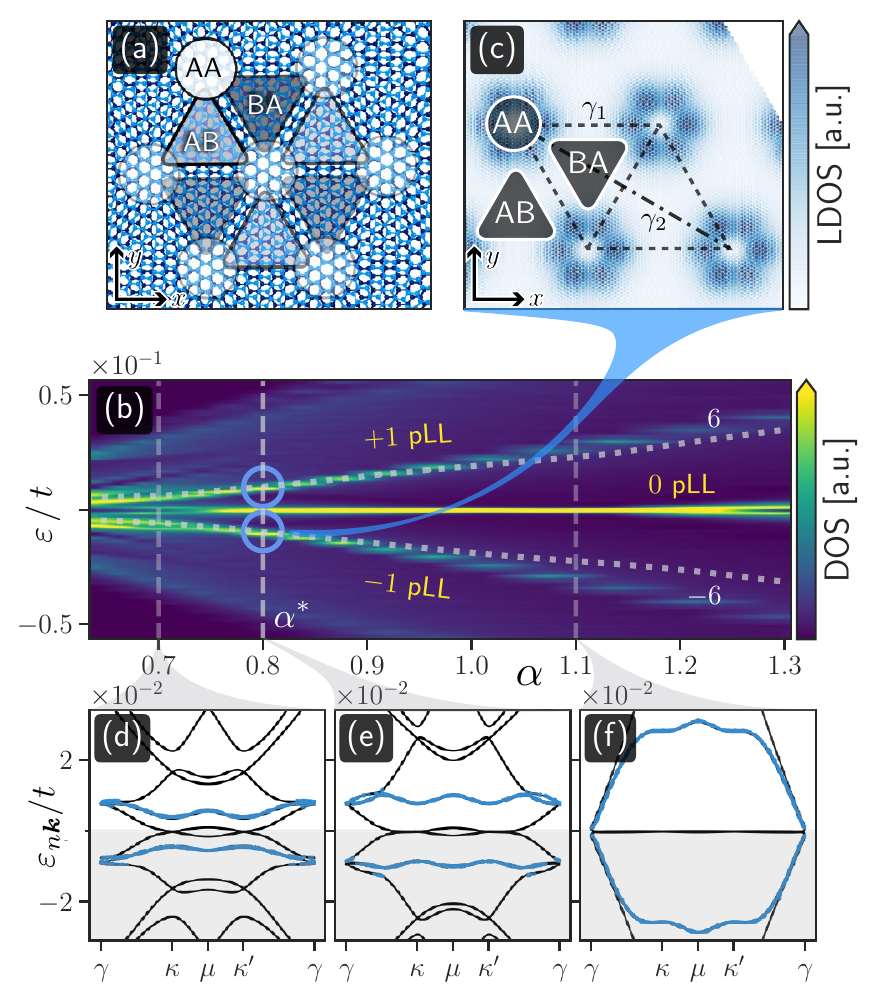}
\caption{\label{fig:fig1}
Tunability of the single-particle bilayer graphene properties as a function of
the twist angle.
    (a)~Stacking two honeycomb lattices (\textit{light/dark blue}) with small
    twist angle $\alpha$ leads to a hexagonal moir\'{e} superlattice. The
    long-range structure has a moir\'e unit cell (\textit{opaque overlay}) with
    varying local stacking order AA/AB/BA, where A and B correspond to the two
    atomic sites in each of the stacked graphene unit cells.
    (b)~Density of states (DOS) as a function of twist angle $\alpha$, showing
    peaks ascribed to emergent flat bands associated with three main
    pseudo-Landau levels (PLLs).  Grey dotted lines denote the position of
    Fermi levels at half-filled $\pm 1$-PLL bands: The two $0$-PLL bands near
    charge neutrality hold $4$ electrons each; similarly, each of the $\pm
    1$-PLL bands holds $4$ electrons and thus is half-filled at a doping level
    of $\pm 6$ electrons per moir\'e cell. At the flat-band angle $\alpha^*=
    0.8^\circ$, the bandwidth of the $\pm 1$-PLLs becomes minimal.
    (c)~Local density of states (LDOS) in the $-1$-PLL, showing the
    emergence of an effective triangular lattice of states localized around
    the AA regions. Dashed lines mark effective hopping amplitudes in an
    effective triangular superlattice Hamiltonian, see text.
    (d)--(f)~Band structures for $\alpha < \alpha^*$, $\alpha = \alpha^*$ and
    $\alpha>\alpha^*$ illustrating how the $\pm 1$-PLL bands evolve to
    generate flat bands at $\alpha^*$ (marked in \textit{blue}). (f) The
    $0$-PLL flattens at the magic angle~\cite{PhysRevB.82.121407,
    Bistritzer2011}. Note that there is a gap to the $\pm 1$-PLLs (small in
    this plotted scale).
}
\end{figure}

Stacking graphene layers at a finite relative angle $\alpha$ produces a moir\'e
superlattice [see Fig.~\ref{fig:fig1}(a)] with properties that are sensitive to
the twist angle~\cite{LopesdosSantos2007, PhysRevB.82.121407, Bistritzer2011}.
The spectrum of the superlattice is composed of graphene bands that are folded
back to the mini-Brillouin zone where they bundle into separate groups; these
can be tuned to become flat at small twist angles $\alpha$.  Such
weakly twisted bilayer graphene then provides a versatile platform to explore
strongly correlated physics. Much work has focused on the so-called magic angle
$\alpha =1.1^\circ$ producing two flat bands near charge neutrality (each
fourfold degenerate) with strong correlations \cite{Bultinck2019, Zhang2019,
Cao2018, Cao2019, 2019arXiv190410153J} and superconductivity
\cite{Cao2018super, Yankowitz2019, Lu2019arxiv} appearing under weak doping.
The question arises, whether other angles and bands can be used in engineering
novel properties. Here, we show that tuning the twist to the angle $\alpha^\ast
\approx 0.8^\circ$ flattens the bands above and below the ones near charge
neutrality. We find that doping these bands to half-filling with $\pm 6$
electrons per triangular supercell produces a correlated state with
ferromagnetic order.

Such flat bands have been termed `pseudo-Landau levels' (PLLs)
\cite{PhysRevB.99.155415, Tarnopolsky2019, 2019arXiv190504515S}; they can be
understood as the result of artificial gauge fields that arise from nonuniform
strain in graphene monolayers~\cite{Vozmediano2010, Low2010, Jiang2017} or
twist-induced strain between layers \cite{San-Jose2012}. Proper tuning of such
strain leads to hopping phases that interfere destructively, localizing the
states and producing flat bands~\cite{PhysRevB.99.155415, Tarnopolsky2019}.
Similar flat-band modes have been proposed in other twisted multilayer
systems \cite{Liu2019arxiv}, such as tiny-angle graphene
bilayer~\cite{RamiresLado2018}, graphene trilayers \cite{2019arXiv190500622M},
graphene bi-bilayers \cite{2019arXiv190108420R,Zhang2019a}, and dichalcogenide
multilayers \cite{PhysRevLett.121.266401}. 

An alternative way to engineer the electronic states in multilayers of Dirac
materials is to apply an interlayer bias~$V$ to induce valley Berry
curvature~\cite{Castro2007}. The latter results from breaking inversion
symmetry, that induces topology through (compensating)
valley fluxes~\cite{San-Jose2013, Rickhaus2018}. Such valley fluxes and
associated valley currents are particularly pronounced at small twist angles
and cause dispersive splittings between bands \cite{San-Jose2013}. This
motivates the idea of engineering the bandwidth and correlations via electrical
bias and geometric twist. Here, we combine twist-induced emergent flat bands
at small twist $\alpha$ and an interlayer bias $V$ in order to manipulate the
strong-correlation physics in a flat-band material with valley-topological
properties.

Flat-band magnetism in twisted bilayer graphene has been discussed before with
a focus on the bands near charge neutrality and manipulation of the magnetic
state by interlayer bias \cite{Gonzalez-Arraga2017, Sboychakov2018,
LopezBezanilla2019} and by pressure \cite{LopezBezanilla2019}.  Rather than
manipulating the $0$th PLL near charge degeneracy, we focus on the $\pm 1$
PLLs that become optimally flat at the smaller angle $\alpha^* \approx
0.8^\circ$ where valley currents are more pronounced. Different from the
$0$th PLL bands that are described by an effective honeycomb lattice and flat
Dirac cones, the $\pm 1$ PLLs are described by a triangular lattice.  We study
the delocalization effect of the bias-induced local valley fluxes and find
that these act against the subtle spatial interference that generates the
original flat bands. The bias-induced dispersion then reduces the correlations
and thus the ferromagnetic (quasi-)order in the material, leading to
electrically tunable magnetism.  We thus arrive at a new platform where
correlations and topology can be controlled at the same time.

Below, we start from the real-space bilayer tight-binding model and find flat
bands at an angle $\alpha^*=0.8^{\circ}$; the states in these $\pm 1$ PLLs
are strongly localized on the triangular superlattice of the moir\'e
structure. We include local interactions on a mean-field level and find the
magnetic instability. In a third step, we include the interlayer bias and map
out the local Berry curvature in real-space that introduces band dispersion
and consequently reduces the magnetic order.

%
Twisted bilayer graphene is formed by stacking two honeycomb lattices of carbon
with a small twist angle $\alpha$, resulting in a moir\'e structure of
(approximate) periodicity $L_M$ that grows inversely with $\alpha$, see
\figref{fig1}(a). This supercell can exceed the lattice constant $a$ in size by
1--2 orders of magnitude and features regions with well-defined local stacking
orders AA, AB and BA, with A and B denoting the inequivalent atomic sites in a
each hexagonal unit cell. We model the twisted bilayer with a real-space
tight-binding Hamiltonian
\begin{align}   \label{eq:ham0}
    H_{0} &  = \sum_{\langle i,j\rangle,s} t\, c^{\dagger}_{i,s} c^{\pdagger}_{j,s}
          +  \sum_{i,j,s} t^{\scriptscriptstyle \perp}_{ij} \, c^{\dagger}_{i,s}c^{\pdagger}_{j,s}
          - \sum_{i,s}\mu\,c^{\dagger}_{i,s} c^{\pdagger}_{i,s}\,,
\end{align}
where $c_{i,s}^{(\dagger)}$ destroys (creates) an electron at site
${\vec{r}_i=(x_i,y_i,z_i)}$ in one of the layers ($z_i=\pm d/2$) with spin
$s=\uparrow,\downarrow$ and $\mu$ is the chemical potential; $t$~is the
nearest-neighbor hopping within each layer and ${t^{\scriptscriptstyle
\perp}_{ij} = t_{\perp} [(z_{i}-z_{j})^{2}/|\vec{r}_{i} -
\vec{r}_{j}|^{2}] \, e^{-(|\vec{r}_{i}-\vec{r}_{j}|-d)/\ell} }$ is the
twist-angle dependent hopping~\cite{Sboychakov2015} from $\vec{r}_i$ to
$\vec{r}_j$ with amplitude $t_{\perp}\simeq 0.12\,t$, range $\ell\simeq 0.13\,a$,
and the interlayer distance $d\simeq 1.4 a$. We utilize a scaling relation
that brings the low-energy physics of small angles $\alpha$ to larger ones by
appropriately increasing the interlayer hopping amplitude $t_\perp$~\cite{PhysRevLett.114.036601,
Gonzalez-Arraga2017, RamiresLado2018, PhysRevB.98.085144, supmat}. This allows us to
perform our analysis at moir\'e unit cells that are small enough for numerical
treatment.

The twist $\alpha$ effectively creates a nonuniform interlayer hopping with a
corresponding gauge field \cite{San-Jose2012,PhysRevB.92.081406}. This, in
turn, leads to a destructive interference that generates our $\pm 1$ PLL flat
bands that naturally lend themselves for strong-correlation physics. In
\figref{fig1}(b), we present the low-energy density of states (DOS) of the
bilayer as a function of its twist angle. Three main PLL bands are indexed with
$-1, 0, 1$, which become flat at the marked specific angles. In this work, we
are interested in the $\pm1$ bands with states that are localized around the AA
regions of the superlattice, see \figref{fig1}(c). At negative (positive)
energies, as a function of the twist angle $\alpha$, the targeted band evolves
from having a negative (positive) effective mass to a positive (negative) one,
see Figs.~\ref{fig:fig1}(d)-(f).  At our ``flat-band'' angle $\alpha^* \simeq
0.8^\circ$ [\figref{fig1}(e)], we achieve maximal isolation of the $\pm1$ PLL
bands in energy. They hold up to $\pm 4$ electrons per moir\'e unit cell and
doping with $\pm 6$ electrons corresponds to their half-filling. Note that the
two $0$ PLL bands hold $\pm 4$ electrons each and become flat and
nearly-degenerate at the magic angle $\alpha\simeq
1.1^\circ$~\cite{PhysRevB.82.121407, LopesdosSantos2007, Bistritzer2011,
Koshino2018, Po2018, Kang2018, Ahn2018arxiv}.

The $\pm 1$-PLL states arrange in a triangular lattice of ``flower''-shaped
Wannier orbitals centered around the AA regions, see \figref{fig1}(c). The
effective low-energy physics for this triangular superlattice and its
corresponding bands can be described by a single-site triangular-lattice
tight-binding Hamiltonian with particles involving four flavors associated with
spin and valley degrees of freedom and effective $C_{3v}$-symmetric hoppings.
It is the absence of Dirac points in the minibands that enables such a
triangular model that reproduces the band dispersion of the targeted band of
the original bilayer tight-binding Hamiltonian \cite{supmat}. In contrast, the
nearly-flat bands at magic angles necessitate a description in terms of Wannier
orbitals arranged on a \emph{honeycomb} superlattice~\cite{Koshino2018, Po2018,
Kang2018, Ahn2018arxiv, Roy2019}. Our low-energy, four-flavor Hamiltonian is
associated with an (approximate) local $\mathrm{SU}(4)$ symmetry; such a model
has been predicted to display $d+id$ superconductivity close to a Mott
insulating state~\cite{Xu2018, 2019arXiv190500033W}, potentially providing a
playground to realize unconventional interacting states.
%

\begin{figure}[!t] \includegraphics[width=3.5in]{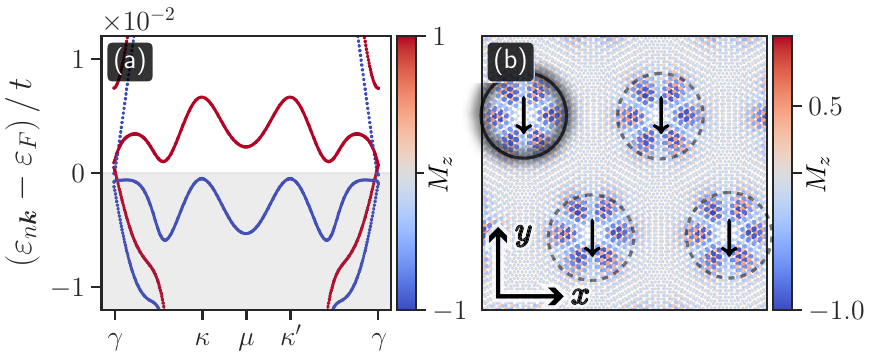}
\caption{\label{fig:fig2}
    Effect of local mean-field interactions [Eq.~\eqref{eq:hubbard} with $U =
    2\, t$] in twisted bilayer graphene at physical twist angle
    $\alpha^*=0.8^\circ$. (a) At $-6$ electron doping, the half-filled flat
    band [cf.~Fig.~\ref{fig:fig1}(e)] exhibits an exchange splitting between
    the up and down spins. (b) The resulting moir\'e structure of local
    magnetic moments. It forms a superlattice of finite collinear moments
    (circled ``$\downarrow$'') in real space. The red (blue) color indicates a
    positive (negative) local expectation value $\expval{S_z}=M_z$ of the spin
    operator that is associated with a net ferromagnetic moment of each AA
    region and a weak local antiferromagnetic texture therein.
    For numerical calculation, we rescaled the parameters to
    $\alpha\simeq 2.86^\circ$, $t_\perp=0.46 t$~\cite{supmat}.
}
\end{figure}

%
We now explore the effect of electronic interactions at $\alpha^\star$ when the
system is doped with $\pm 6$ electrons. We introduce interactions in a minimal
way using a local Hubbard term, yielding the interacting Hamiltonian $H = H_0 +
H_U$ with
\begin{equation} \label{eq:hubbard}
    H_U = U\sum_{i} n_{i\uparrow} n_{i\downarrow},
\end{equation}
where $U$ is the interaction strength and $n_{i,s} = c^{\dagger}_{i,s}
c^{\pdagger}_{i,s}$ is the local density operator at each site. We chose the
interaction strength $U=2\,t$ which accounts for a physical charging
energy and our numerical rescaling, see Refs.\
\cite{supmat,Guinea2018,Cea2019arxiv}. We address this interaction term
self-consistently using the spin-collinear mean-field ansatz $n_{i\uparrow}
n_{i\downarrow} \approx \langle n_{i\uparrow} \rangle n_{i\downarrow} +
n_{i\uparrow} \langle n_{i\downarrow}\rangle - \langle n_{i\uparrow} \rangle
\langle n_{i\uparrow}\rangle$. 
We find that the interaction [Eq.~\eqref{eq:hubbard}] mainly affects the
nearly-flat band of the superlattice, i.e., it induces a Stoner instability
that splits the band, see \figref{fig2}(a). This interaction-induced exchange
splitting is associated with a net magnetic moment of approximately $2\,\mu_B$
per moir\'e unit cell (obtained by integrating over the
mildly-antiferromagnetic texture) aligned ferromagnetically between cells.
Given the macrostructure of the system, this yields a triangular superlattice
of local magnetic moments, see
\figref{fig2}(b).

The emergence of magnetism in this flat-band angle regime is analogous to
observations of magnetic instabilities at other angles and dopings
\cite{Sboychakov2018, Gonzalez-Arraga2017}, in particular, in magic-angle
superlattices~\cite{Thomson2018,2018arXiv181202550S,Kang2018aArxiv}. Unique to
our regime is the sensitivity of the $\pm 1$-PLL bands to an interlayer
voltage-bias: as elaborated below, the latter modifies the electronic spectrum
of the bilayer and hence allows for the tuning of the aforementioned magnetic
order.

\begin{figure*}[!t]
\includegraphics[width=7in]{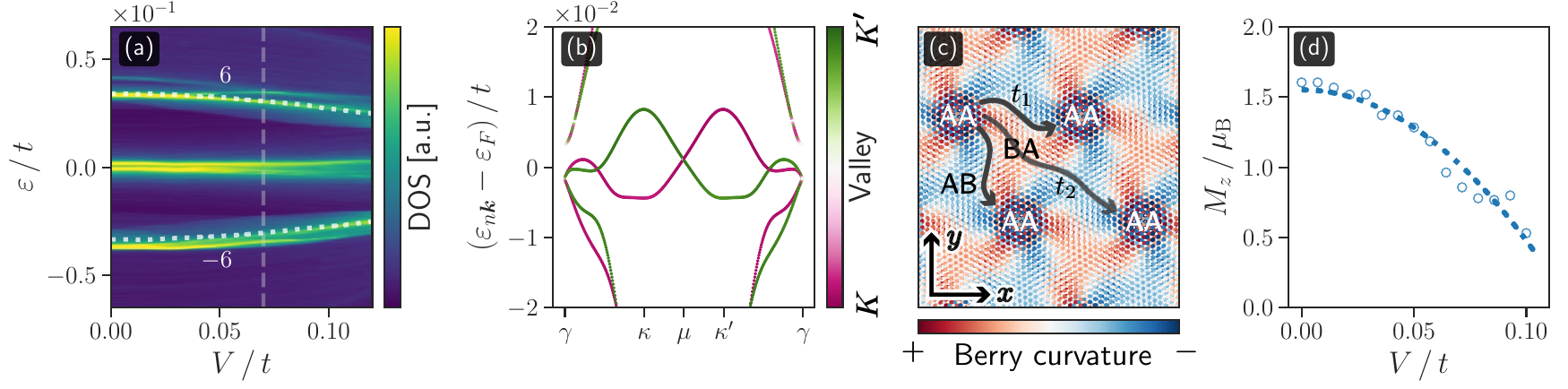}
\caption{\label{fig:fig3}
    Effects of interlayer bias $V$ on the flat bands of bilayer graphene at
    twist angle $\alpha^*=0.8^\circ$, (a)--(c)~without and (d)~with
    interactions.
    (a)~The sharp peaks in the density of states (DOS) broaden and vanish with
    increasing interlayer bias $V$; \textit{dotted lines} mark Fermi energies
    for dopings $\pm6$.
    (b)~At $-6$ electron doping, the half-filled flat bands disperse upon
    introduction of a small interlayer voltage bias [$V/t=0.07$,
    \textit{dashed line} in panel (a)] to form two independent
    valley-polarized bands (\textit{green and purple}) [cf.~Fig.
    \ref{fig:fig1}(e)].
    (c)~The local valley Berry curvature in position space $\partial
    \Omega_\valley(\vec{r})/\partial \omega$ [Eq.~\eqref{eq:localvalleyhall}]
    (integrated over the bands shown in b) can be understood in terms of an
    emergent valley magnetic field in the AB/BA regions. This field destroys the
    fine-tuned interference, inducing a finite bandwidth in the originally flat
    bands.
    (d)~The total interaction-induced [$U=2\,t$] mean-field magnetization
    $M_z$ at the twist angle $\alpha^*$ decreases with interlayer bias $V$. The
    bias-induced dispersion quenches the interactions and the formation of
    magnetic order.
    For numerical calculation, we rescaled the physical parameters to
    $\alpha\simeq 2.86^\circ$, $t_\perp=0.46 t$~\cite{supmat}.
}
\end{figure*}

%
We now discuss the implications of an interlayer voltage-bias as described by
the Hamiltonian $H=H_0 + H_V$ (at $U=0$) with
\begin{equation} \label{eq:hamV}
    H_V = \frac{V}{2} \, \sum_{i,s} \mathrm{sgn}(z_i)\, c^{\dagger}_{i,s} c^{\pdagger}_{i,s} ,
\end{equation}
where $V$ is the interlayer bias and $\mathrm{sgn}(z_i) = \pm1$ for the top and
bottom layers, respectively. This bias lowers the symmetry from $C_{3v}$ to
$C_3$ by breaking the inversion symmetry between the layers. Finite $V$
broadens and eventually washes out the singularity in the density of states
[\figref{fig3}(a)] by introducing a dispersive valley-splitting into the
band structure [\figref{fig3}(b)]. The substantial impact of the bias $V$ on
these flat bands is in stark contrast to the magic-angle regime, where it was
found that a small bias has a negligible effect~\cite{Yankowitz2019}.

%
To better understand the mechanism that lifts the valley degeneracy, we analyze
the spatial profile of the frequency-resolved valley Berry
curvature~\cite{Chen2011}
\begin{equation} \label{eq:localvalleyhall}
    \frac{\partial \Omega_\valley(\vec{r})}{\partial \omega}
    =
\int_{\scriptscriptstyle\textrm{BZ}} \frac{d^2 k}{(2\pi)^2} \,\frac{\epsilon_{\alpha \beta}}{2}
\langle \vec{r} | G_{\valley} (\partial_{k_\alpha}G_{\valley}^{-1}) (\partial_{k_\beta}G_{\valley}) | \vec{r} \rangle,
\end{equation}
where $\int_{\scriptscriptstyle \textrm{BZ}} d^2 k \cdots$  denotes
integration over the Brillouin zone and $\epsilon_{\alpha\beta}$ is the
Levi-Civita tensor.  We denote $G_\valley = [\omega - H(\vec k)+i0^+]^{-1}
\mathcal{P}_\valley$ as the Green's function of the Bloch Hamiltonian $H(\vec
k)=H_0(\vec k)+H_V(\vec k)$, and $\mathcal{P}_\valley$ is the valley
polarization operator~\cite{Colomes2018, RamiresLado2018,2019arXiv190205862R}
that weighs the states with $\pm 1$ depending on which graphene valley ($K,K'$)
they originate from. This allows us to find the valley-Chern number through
integration over energies and over the unit cell, i.e., $C_\valley =
\int_{\scriptscriptstyle\textrm{UC}} d^2r \int_{-\infty}^{\mu} d\omega  \;
\partial_\omega \Omega_\valley(\vec{r})$, which only takes nonzero values if
time-reversal or inversion symmetry are broken. By breaking inversion
symmetry, the interlayer bias~$V$ produces a finite valley-Chern number with
the spatial texture in the Berry curvature as shown in \figref{fig3}(c),
featuring alternating signs between AB/BA-stacked regions~\cite{Castro2007,
Zhang2013, Tong2016}. These, in turn, can be interpreted as a fake staggered
valley magnetic field. At smaller angles $\alpha < \alpha^*$, this
spatial texture in the Berry curvature sharpens up and eventually leads to the
valley helical networks~\cite{San-Jose2013} recently observed in
experiments~\cite{Rickhaus2018}.

%
If we regard the formation of flat bands in \figref{fig1}(e) as the result of
fine-tuned destructive interference between neighboring Wannier
orbitals~\cite{PhysRevB.82.121407,Bistritzer2011}, we can explain how the
dispersive valley-splitting in \figref{fig3}(b) arises: the emergent
valley-chiral magnetic field induced by the interlayer bias \emph{alters} the
relative phases between the Wannier orbitals, and creates additional
valley-splittings. This effect can be captured by the effective triangular
superlattice Hamiltonian via the inclusion of Peierls phases in the
hoppings~\cite{supmat}. These phases correspond to a finite magnetic flux per
triangle that averages to zero over one unit cell, see \figref{fig3}(c),
reminiscent of Haldane's model for Chern insulators~\cite{PhysRevLett.61.2015}.
As the two valleys are time-reversal-symmetric partners these fluxes have
opposite sign, thus canceling one another locally.  To summarize, the
interlayer bias in the full model takes the role of the valley-magnetic flux
per plaquette in the low-energy model.

%
We can now analyze how the interlayer bias affects the interaction-induced
correlated state by considering the Hamiltonian $ H = H_0 + H_U + H_V $. We
expect that the interlayer-bias-induced band dispersion modifies the magnetic
state discussed in \figref{fig2}. Indeed, as shown in \figref{fig3}(d), the
ground state magnetization is substantially quenched even for moderate bias
($V\simeq 100 \textrm{ meV}$). Hence, the interlayer bias serves as an external
control for the correlated state at our \textit{flat-band}-angle. We once more
emphasize that this feature is in striking contrast to magic-angle graphene,
where a small interlayer bias does not substantially change the correlated
state~\cite{Yankowitz2019}.


There are three interesting avenues for further investigations beyond the scope
of this work. First, we emphasize that our analysis does not take into account
possible lattice relaxations~\cite{PhysRevB.96.075311, Lucignano2019}, which
may impact the specific angle at which the targeted band becomes sufficiently
flat. Second, the existence of twist-angle disorder~\cite{Beechem2014} is
likely to modify the width of the targeted band and uniform twist angles may be
required in order to observe the correlated state -- similarly to the
phenomenology shown at the magic angle~\cite{Lu2019arxiv}.
Third, the triangular nature of the superlattice suggests that a spin-spiral
state may exist that is energetically competitive with the ferromagnetic
configuration discussed above. Indeed, we performed our self-consistent
mean-field analysis for several candidates and found indications for a
$120^\circ$ spin-spiral state (maximally antiferromagnetic) that competes with
the collinear order. This observation leads us to conclude that the system is
profoundly frustrated, with sizable antiferromagnetic exchange coupling to
second-neighbor moir\'e cells. Since such triangular lattices have been
proposed to give rise to spin-liquid phases~\cite{Kaneko2014}, further
investigations into our proposed flat-band system might unveil a nontrivial
realization of such physics.

To summarize, we have shown that electronic correlations can arise in twisted
graphene bilayers at fillings of 6 electrons/holes per moir\'e unit cell. This
correlated regime is shown to appear at angles around $0.8^\circ$ at doping
levels of $\pm 6$ electrons per moir\'e unit cell, for which the chemical
potential falls into one of the $\pm 1$-pseudo-Landau levels. For that regime,
we show that interactions promote the formation of local magnetic moments in
the moir\'e supercells arranging on a triangular lattice.  Furthermore, we have
shown that the interlayer bias can be used to control the magnetic instability.
The origin of this tunability was demonstrated to be related to the control
over an effective staggered valley-magnetic field in the heterostructure, that
modifies fine-tuned interferences in the superlattice states, thereby
substantially affecting the low-energy dispersion. Our results put forward a
new regime in twisted graphene multilayers hosting correlations that result in
a magnetic instability that is highly tunable with weak interlayer biases.


\begin{acknowledgments}
We acknowledge financial support from the Swiss National Science Foundation.
J.~L.~L. acknowledges financial support from the ETH Fellowship program.
\end{acknowledgments}



\end{document}


\relscale{1.06}

\title{Supplementary material for\\ ``Electrically Tunable Flat Bands and Magnetism in Twisted Bilayer Graphene''}

\author{T. M. R. Wolf}
\author{J. L. Lado}
\author{G. Blatter}
\author{O. Zilberberg}
\affiliation{Institute for Theoretical Physics, ETH Zurich, 8093 Zurich, Switzerland}

\date{\today}

\begin{abstract}
In this supplemental material, we provide additional details on (i) methods
used for the analysis of the microscopic model of the twisted bilayer
(section~\ref{sec:micro}) and (ii) the effective (phenomenological) low-energy
model describing the triangular lattice of localized AA Wannier orbital, as
well as its effectiveness in capturing the effects observed in the main text
(section~\ref{sec:super}). Specifically, using (ii), we discuss how the
emergence of flat bands $\pm 1$-PLLs can be understood to be a result of
destructive interference (i.e., suppressed hopping amplitudes) and  how the
interlayer bias (interpreted as Haldane-like valley-flux texture in the moir\'e
supercell) detunes the interference and induces valley-dependent dispersion.
\end{abstract}

\maketitle

\section{Methods for analyzing the microscopic tight-binding model}
\label{sec:micro}

In the main text, we model the non-interacting twisted bilayer with a
real-space tight-binding Hamiltonian [see Eqs.~(1) and (3) in the main text]. 
%
%
The spectrum of the twisted bilayer system results from backfolding and
hybridizing the graphene bands of both layers in a mini-Brillouin zone. In the
current work, the spectrum is obtained using exact diagonalization. Due to
computational resource bounds, this exact approach limits how large we can
choose our moir\'e unit cells. To reduce the size of the unit cells (and hence
the number of atoms within them), we use the property that the low-energy bands
at small angles only depend on the ratio
$t_\perp/\sin(\alpha/2)$~\cite{PhysRevLett.114.036601,Gonzalez-Arraga2017,RamiresLado2018,PhysRevB.98.085144},
i.e., we can bring the low-energy physics of smaller angles to larger ones by
increasing the interlayer coupling $t_\perp$, or more precisely:
$\alpha\to\alpha'$ with $t_{\perp} \to t_{\perp}
\sin(\alpha'/2)/\sin(\alpha/2)$ and $t \to t\, \sin(\alpha'/2)$.

To produce Fig.~1, we use the parameters $t_{\perp}=0.23 t$, $V=0$ and twist
angles $\alpha_{n,1}\approx 5.1 ^\circ,\dots,1.12^\circ$ with $n=6,\dots,29$.
Making use of the scaling transformation, this corresponds to the physically
realistic parameters $t_\perp'=0.12 t$ with $\alpha'$ between $0.57^\circ$ and
$1.12^\circ$. To produce Fig.~2, we use $t_\perp=0.46 t$, $V=0$ and
$\alpha_{11,1}\simeq 2.86^\circ$, which is the rescaled equivalent of
$t_\perp'=0.12 t$ for twist angle $\alpha'=0.8^\circ$. To produce Fig.~3, we
use the same parameters as in Fig.~2, but in (a),(d) we vary $V=0,\dots0.12 t$,
in (b),(c) $V=0.07 t$.


\section{Effective triangular tight-binding model describing the $\pm 1$-PLLs} \label{sec:super}

\begin{figure*}[!b]
  \centering
  \includegraphics[width=5in]{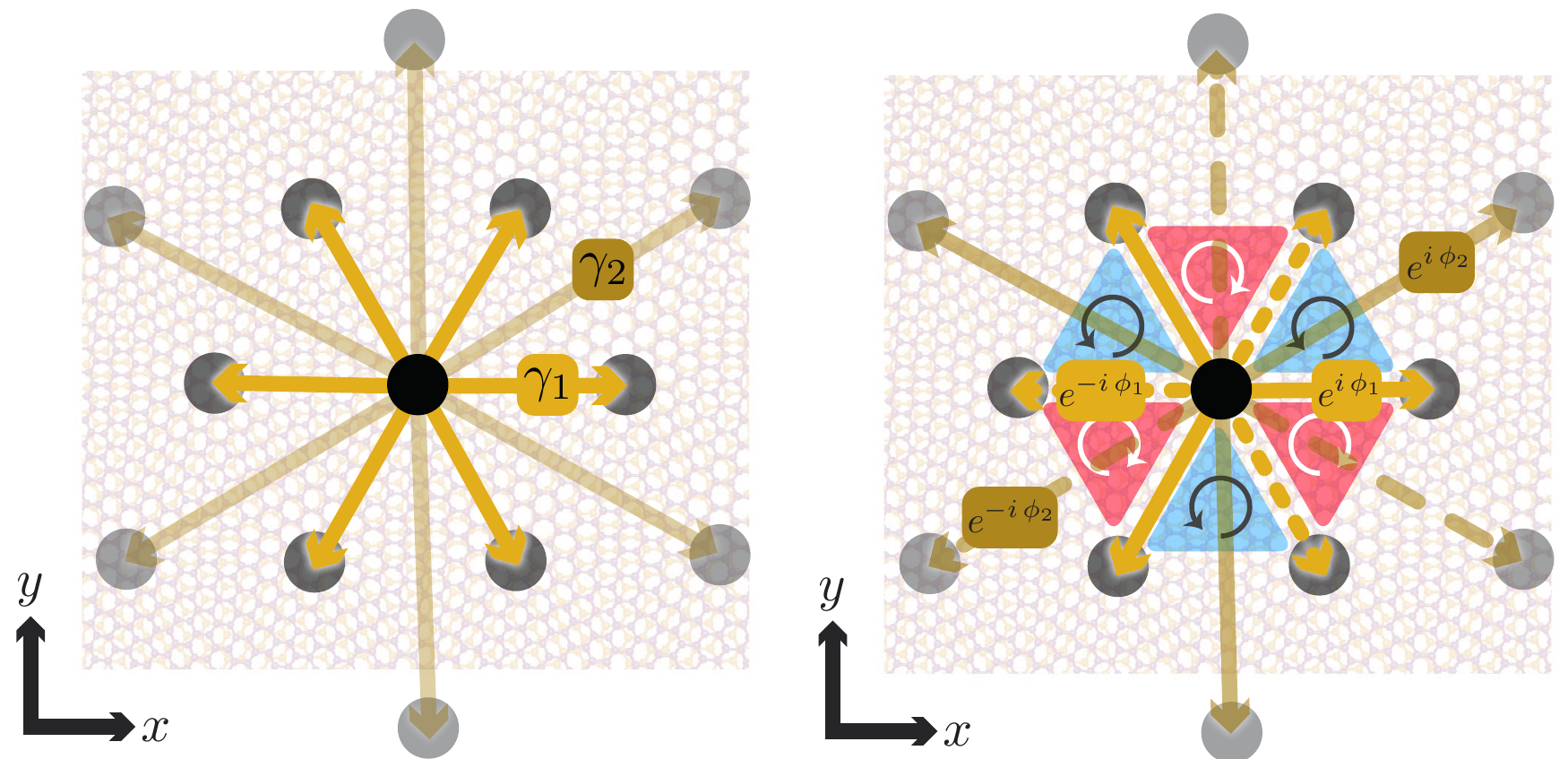}
  \caption{\label{fig:sfig1} Sketch of the effective superlattice for Wannier
  states localized near AA stacked regions (black circles) in the moir\'e
  structure (a) without interlayer bias and (b) and with finite interlayer bias
  $V>0$. The effective physics of the $\pm 1$ PLL bands is captured by nearest-
  and next-nearest-neighbor hopping amplitudes $\gamma_{1}$ and $\gamma_{2}$.
  The interlayer bias induces complex phases in $\gamma_{1}$ and $\gamma_{2}$
  that encode the emergent valley magnetic flux texture piercing the triangular
  unit cells (red/$\circlearrowright$,  and blue/$\circlearrowleft$ in- and
  out-of-plane flux, respectively).
  }
\end{figure*}

The texture of the local density of states for the $\pm 1$~PLL bands
[cf.~Fig.~1(c)] suggests that the eigenstates within these bands form
Wannier-orbitals around the AA regions that are coupled on a triangular
lattice. Furthermore, these Wannier orbitals are spin-degenerate and
independently result from both valley degrees of freedom of the bilayer, see
Fig.~\ref{fig:sfig1}(a). We can, therefore, coarse grain these orbitals to
effective atomic sites and describe their coupling using a tight-binding model
\begin{align} \label{eq:lowh}
   H_{{\rm eff}} = \sum_{\langle\eta\eta'\rangle,\sigma} \gamma_1 \;
   d^{\dagger}_{\eta\sigma}  d^{\pdagger}_{\eta'\sigma} + \sum_{\langle\langle\eta\eta'\rangle\rangle,\sigma} \gamma_2 \;
   d^{\dagger}_{\eta\sigma}  d^{\pdagger}_{\eta'\sigma},
\end{align}
where $\sigma=(s,\nu)$ labels the four orbitals (spin $s=\uparrow,\downarrow$
and valley $\nu=\pm$), and $d_{\eta\sigma}^{(\dagger)}$ destroys (creates) an
electron in the respective  Wannier orbital in AA region $\eta$.  $\gamma_{1}$
and $\gamma_{2}$ are real nearest and next-nearest hopping amplitudes
respecting $C_{3v}$ symmetry.

\subsection{Flat band through destructive interference}

\begin{figure*}[!b]
  \centering
	\includegraphics[width=5.8in]{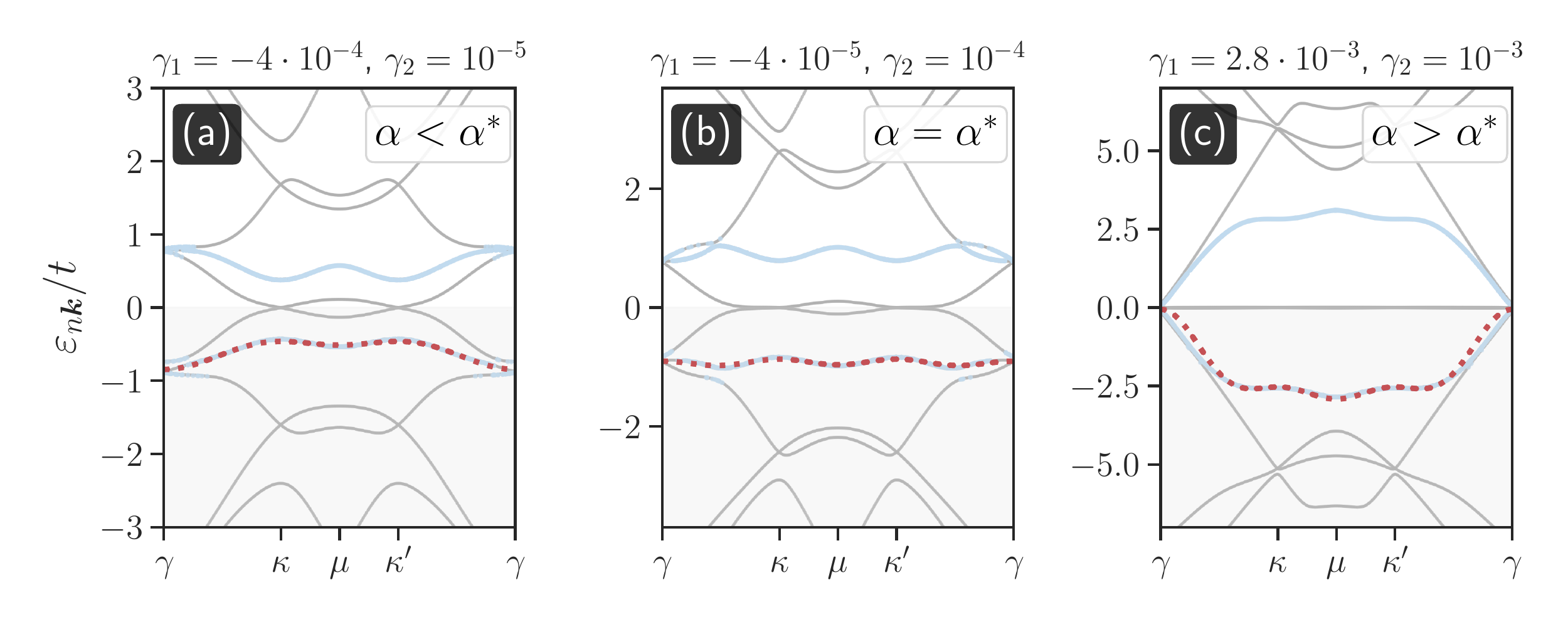}
	\caption{\label{fig:sfig2}
  Band dispersion of the effective model \eqref{eq:lowh} \emph{(dotted red)}
  such that it fits the $-1$-PLL band dispersion of the twisted bilayer
  graphene \emph{(pale blue)} for different twist angles (a) $\alpha<\alpha^*$,
  (b) $\alpha<\alpha^*$, and (c) $\alpha>\alpha^*$ [cf. Figs.~1(d)--(f)]. We
  see that the effective hopping parameter $\gamma_{1}$ of the dispersive band
  in (a) and (c) switches  sign while $\gamma_2$ remains approximately
  constant. This, in turn, generates the negative and positive effective mass
  of the band. At the transition (b), $\gamma_1\sim 0$ and the band flattens. 
	}
\end{figure*}

Generally, the effective hopping in the low-energy model \eqref{eq:lowh} is
controlled by the twist angle $\alpha$ and the interlayer hopping $t_\perp$ in
the microscopic model, i.e., $\gamma_{i}\equiv\gamma_{i}(\alpha, t_\perp)$.  In
Fig.~\ref{fig:sfig2}, we fit the effective hopping amplitudes $\gamma_1$ and
$\gamma_2$ such that we can phenomenologically reproduce the $-1$-PLL band
dispersion in Figs.~1(d)--(f).

\subsection{Interlayer voltage bias as valley-flux}

As discussed in the main text, a voltage-bias between the twisted graphene
layers breaks inversion symmetry and consequently lifts the degeneracy between
the two valleys of the material. Our analysis of the valley Berry curvature
revealed a fake valley (or `chiral') magnetic field that we can incorporate
into our single-site four-orbital triangular hopping model \eqref{eq:lowh} via
a Peierl's substitution with opposing fluxes between the two valleys, i.e.,
\begin{align}
\gamma_{j} \mapsto \gamma_{j} \, e^{i\,\nu\,2\pi\,\phi_{j}},
\label{complex_hops}
\end{align}
where $\nu=\pm$ is again the valley degree of freedom. The phases should be compatible with the Haldane-like valley flux pattern in Fig.~3(c), i.e., $0<\phi_1<1$ and $0\leq\phi_2\ll 1$.

\begin{figure*}[!t]
  \includegraphics[width=4.8in]{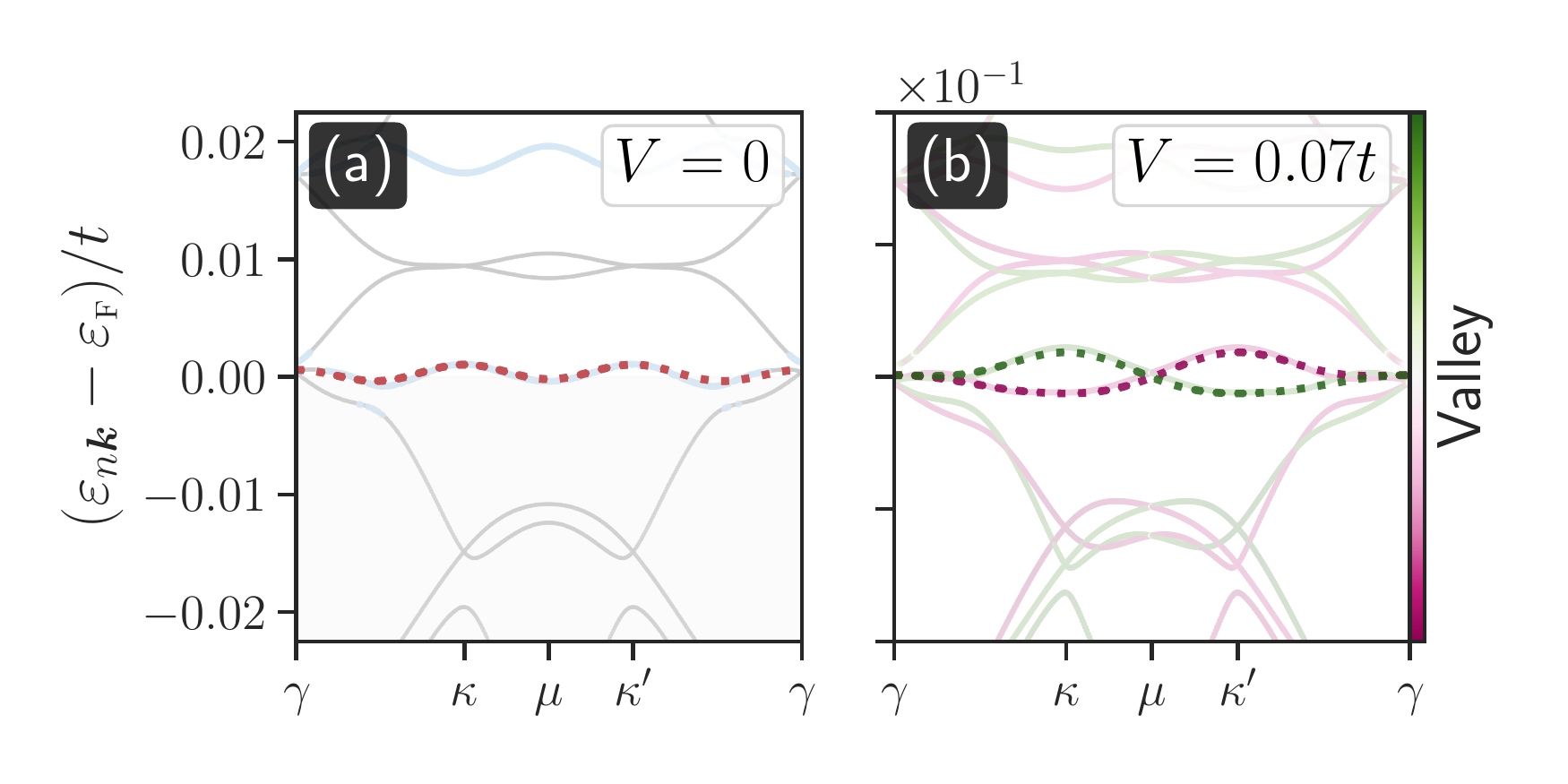}
  \caption{\label{fig:sfig3}
  Band dispersion of the effective triangular model \eqref{eq:lowh} once
  complex hopping amplitudes \eqref{complex_hops} have been incorporated in
  order to account for the valley-flux. The effective band dispersion was
  fitted to that of the $-1$-PLL band of twisted bilayer graphene \emph{(pale
  colors)} for different voltage-bias (a) $V=0$ and (b) $V=0.07t$ [cf.
  Fig.~3(b)]. As discussed in the main text, the emergent valley-magnetic
  fluxes (that are induced by the interlayer bias) lift the (approximate)
  valley degeneracy.
  }
\end{figure*}

In Fig.~\ref{fig:sfig3}, we show that by selecting the appropriate hopping parameters $\gamma_{j}$ \emph{and} phases $\phi_{j}$ reproduces the interlayer bias-induced valley splitting. The parameters of panel (a) are the same is in Fig.~\ref{fig:sfig2}(b) and those for panel (b) 
have phases $\phi_1$ = $0.75$ and $\phi_2=0.1$.

\section{Electronic interactions in twisted multilayer graphene}

\subsection{Hubbard versus screened Coulomb interactions}\label{sec:HubbardvsCoulomb}

In twisted graphene multilayers, the long-ranged part of the Coulomb
interaction is known to enhance the correlated gaps that emerge in flat-band
regimes~\cite{Guinea2018,Cea2019arxiv}.  In what follows, we show this
enhancement by comparing the band dispersion of the (self-consistent)
mean-field Hamiltonians for (a) a local Hubbard interaction with (b) that of
(screened) Coulomb interaction. 
To that end, let us consider the more generic interaction
\begin{equation} \label{Vcou}
    H_U = \sum_{i,j} V(|\vec r_i - \vec r_j |) \; n_{i\uparrow} \; n_{j\downarrow},
\end{equation}
where $n_{i,s} = c^{\dagger}_{i,s} c^{\pdagger}_{i,s}$ is the local density
operator at each positions $\vec{r}_i$ and $V(r)$ is the two-particle
interaction. The latter shall contain two contributions: an onsite (Hubbard)
repulsion $U_0$ and a long-ranged Coulomb potential with screening length
$\lambda^{-1}$, i.e.,
\begin{equation} \label{nlcou}
    V(r) = \begin{cases}
              U_0, & \text{for} \quad r = 0 \\
              \frac{V_0}{r} e^{-\lambda (r-a)} & \text{for} \quad  r \ne 0 ,
            \end{cases}
\end{equation}
where $a$ the carbon-carbon distance. Note that $V_0=0$ recovers the Hubbard
Hamiltonian used in the main text.

\begin{figure*}[!t]
\centering
\includegraphics[width=4.5in]{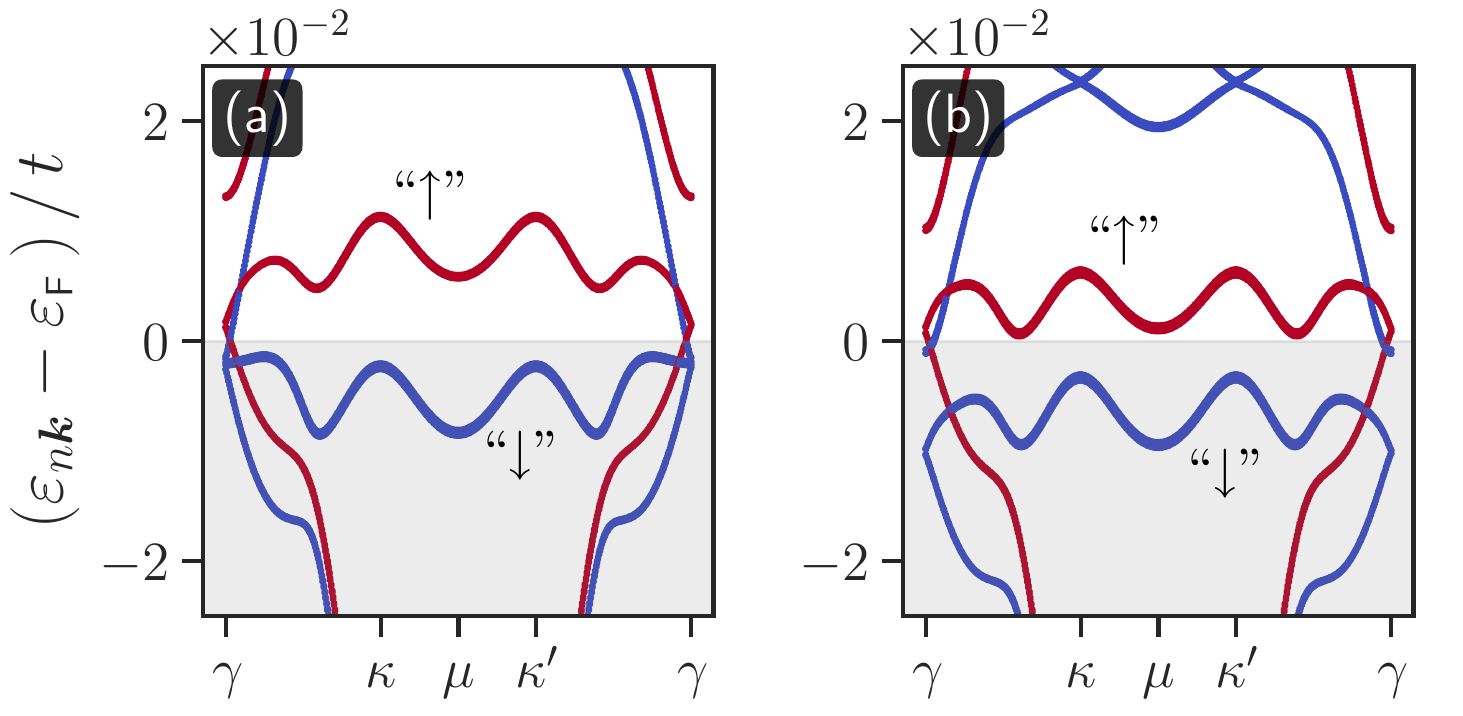}
\caption{\label{fig:sfig4}
  Band dispersion in presence of electronic interactions [see Eq.~\eqref{Vcou}]
  for (a) onsite Hubbard repulsion $U_0=2t$ ($V_0=0$) and for (b) reduced
  Hubbard repulsion $U_0=0.5t$ but in presence of screened Coulomb interactions
  ($V_0=0.37t$ and $\lambda^{-1}=a$, where $a$ is the inter-atom distance);
  obtained from self-consistent mean-field iteration. Near the chemical
  potential, the two band dispersions are qualitatively very similar, showing
  that the long-ranged part of the interaction effectively renormalizes the
  onsite Hubbard repulsion. Note that we used a twist angle $\alpha_{n,1}$ with
  $n=8$.
}
\end{figure*}

As in the main text, we perform a mean-field decoupling in Eq.~\eqref{Vcou},
i.e., $n_{i\uparrow} n_{i\downarrow} \approx \langle n_{i\uparrow} \rangle
n_{i\downarrow} + n_{i\uparrow} \langle n_{i\downarrow}\rangle - \langle
n_{i\uparrow} \rangle \langle n_{i\uparrow}\rangle$. We again iterate
self-consistently to obtain the ground state of the mean-field Hamiltonian
$H_0+H_V$. Fig.~\ref{fig:sfig4} shows the self-consistent bandstructure for
\emph{(a)} the pure Hubbard case ($V_0=0$, $U_0=2t$) that we considered in the
main text, and \emph{(b)} the screened Coulomb case with a reduced onsite
repulsion $U_0=0.5t$ and $V_0=0.37t$. Both systems show a similar exchange
splitting, despite a significantly smaller local interaction $U_0$ in case (b).
We conclude that the effect of a (more realistic) screened Coulomb interaction
on the minibands near the chemical potential can be qualitatively reproduced by
renormalizing (enhancing) the onsite repulsion and discarding the long-ranged
term in the Hamiltonian.

\subsection{Rescaling the interaction strength along with the unit cell}

As explained above (see section \ref{sec:micro}), we rescaled our unit cells
and interlayer hopping amplitudes towards larger angles, i.e.
$\alpha\to\alpha'=\lambda \,\alpha$ with $t_\perp\to t_\perp'=\lambda \,
t_\perp$. In what follows, we deduce the scaling of the interaction strength
$U_0$ from the charging energy $U_{\rm eff}=U_0/N$ of the low-energy states
\cite{Guinea2018}. Here, $N$ is the number of atoms per moir\'e cell and grows
inversely with $\alpha^2$ (it is proportional to the area of the unit cell). To
keep the low-energy physics unchanged, we have to keep the ratio of typical
bandwidths and charging energy fixed when performing the rescaling:
\begin{align}
  U_0/t_\perp=U_0'/t_\perp' && \text{for} && \alpha\to\alpha'=\lambda\, \alpha, && t_\perp\to t_\perp'=\lambda\,  t_\perp, && N \to N'= \lambda^{-2}\, N.
\end{align}
This yields the rescaled Hubbard interaction $U_0'=\lambda^{-1}\,U_0$.

For the parameters of our system, this would imply using a value $U_0 \approx
0.6 t$ for $n=11$ for the realistic interaction term of Eq.~\eqref{Vcou},
together with the long-range part of the Coulomb interaction. We note that
performing non-local mean-field calculations is substantially more
computationally demanding than local ones. For this reason, in the main
manuscript we have captured the effect of the non-local term by a renormalized
local Hubbard Hamiltonian, namely taking $U_0 = 2t$ and $V=0$ in
Eq.~\eqref{Vcou}, as shown in Fig. \ref{fig:sfig4}(a)  (see section
\ref{sec:HubbardvsCoulomb}). This choice of parameters yields a physical
charging energy of $30$ meV at $\alpha\simeq 0.8^\circ$, similar to charging
energies at the magic angle.


\bibliography{biblio}
